\documentclass{emulateapj}

\newcommand{\ltsim}{\protect\raisebox{-0.5ex}{$\:\stackrel{\textstyle <}
        {\sim}\:$}}
\newcommand{\gtsim}{\protect\raisebox{-0.5ex}{$\:\stackrel{\textstyle >}
        {\sim}\:$}}

\newcommand{\msun}{M_{\odot}}
\newcommand{\lsun}{L_{\odot}}
\newcommand{\rsun}{R_{\odot}}

\begin{document}

\title{Mass Transfer in Close, Rapidly Accreting Protobinaries:\\An Origin for Massive Twins?}

%\centerline{DRAFT: \today}
\slugcomment{Accepted to the Astrophysical Journal, February 19, 2007}

\author{Mark R. Krumholz\footnote{Hubble Fellow} and Todd A. Thompson\footnote{Lyman Spitzer Jr.~Fellow}}
\affil{Department of Astrophysical Sciences, Princeton University,
Princeton, NJ 08544}
\email{krumholz@astro.princeton.edu, thomp@astro.princeton.edu}

\begin{abstract}
Rapidly accreting massive protostars undergo a phase of deuterium shell
burning during pre-main sequence evolution that causes them to swell to tenths of an
AU in radius. During this phase, those with close binary companions
will overflow their Roche lobes and begin transferring mass. Since
massive stars frequently have companions at distances well under 1
AU, this process may affect the early evolution of a substantial
fraction of massive stars. We use a simple protostellar evolution
model to determine the range in accretion rates, mass ratios, and
orbital separations for which mass transfer will occur, and we compute
approximately the stability and final outcome of the transfer
process. We discuss how mass transfer affects the demographics of
massive binaries, and show that it provides a natural explanation for
the heretofore unexplained population of massive ``twins'', high mass
binaries with mass ratios very close to unity.
\end{abstract}


\section{Introduction}

Massive stars form in regions of high pressure and density that
produce high accretion rates. Simple order-of-magnitude estimates for
the properties of observed massive protostellar cores suggest that
accretion rates of $\sim 10^{-4} - 10^{-2}$ $\msun$ yr$^{-1}$ are
typical \citep{krumholz06b}, and both detailed analytic models
\citep{mckee03} and numerical simulations \citep{banerjee07a, krumholz07a} produce
accretion rates in this range. Observations of protostellar outflows
with mass fluxes up to $\sim 10^{-2}$ $\msun$ yr$^{-1}$
from luminous embedded protostars \citep{henning00} also suggest high
accretion rates. 

The combination of rapid accretion and high mass produces protostars with very large radii, for two
reasons. First, \citet{stahler88} shows that prior to the onset of deuterium burning, the radius of an accreting protostar is determined primarily by the specific entropy of the gas after it passes through the accretion shock, radiates and settles onto the surface, and is eventually buried in enough optical depth to prevent it from radiating further. A high accretion rate reduces the amount of time an accreted gas element has to radiate before it is buried, and thus produces higher specific entropy and larger radius. Second, \citet{palla91, palla92} show that after deuterium burning begins, massive protostars pass through a period of shell burning. This occurs because, as contraction raises the core temperature, the opacity decreases to the point where the core becomes convectively stable. Accreting deuterium cannot penetrate the radiative core and, as a result, deuterium burning occurs in a shell around it. In a process analogous to that in a red giant, this produces rapid expansion of the envelope above the burning layer. Together, these two effects can produce radii of several tenths of an AU during pre-main sequence evolution of rapidly accreting stars.

Large radii create the possibility for mass transfer in close binaries. Most massive stars appear to be in multiple systems \citep[e.g.][]{preibisch01, shatsky02, lada06}. While the semi-major axis distribution of massive stars is not well-determined, due to low statistics and complex selection biases, there appears to be a significant population of very close binaries. The WR20a system, the most massive binary known, has a separation of only $0.25$ AU \citep{bonanos04, rauw05}. The semi-major axis of the massive detached eclipsing binary D33J013346.2+304439.9 in M33 is 0.22 AU \citep{bonanos06}. \citet{harries03} and \citet{hilditch05} report a sample of 50 OB star eclipsing binaries with periods of 5 days or less taken from the Optical Gravitational Lens Experiment survey \citep[OGLE;][]{udalski98} of the SMC. This period limit corresponds to a semi-major axis of $0.26 M_{100}^{1/3}$ AU, where $M_{100}$ is the total mass in units of $100$ $\msun$. Roughly half the systems are detached binaries, indicating that they have not yet undergone any post-main sequence mass transfer. It is therefore likely that these systems formed at close to their current separations, or possibly even closer, since conservation of energy dictates that mass loss from winds during main sequence evolution widens the orbits of massive binaries. This means that systems such as WR20a, DD33J013346.2+304439.9, and the SMC binares almost certainly experienced a phase of mass transfer during their pre-main sequence evolution.

To study how this process will affect massive binaries, we proceed as follows. We assume that a ``seed" close protobinary system forms while the system is still embedded and accreting. Such a binary may form via one of two possible mechanisms that have been proposed: direct fragmentation of a massive molecular cloud core and subsequent capture of two protostars into a tight binary \citep[e.g.][]{bonnell05}, or fragmentation of a disk around a massive protostar and subsequent migration of a stellar-mass fragment inward \citep[e.g.][]{kratter06,krumholz07a}. Although these two mechanisms arise in the context of specific star formation models, we emphasize that seed binaries of this sort are expected to form early in massive protostars' lives in the context of any model where massive protostars form by accretion \citep[e.g.][]{mckee03, keto06b, krumholz07a}. Moreover, recent observations by \citet{apai07} show that massive stars have a high intrinsic binary fraction even while they are still embedded, providing observational support for this hypothesis. It is worth noting, however, that stable binaries may not form until much later in massive protostellar evolution if massive stars form by collisions \citep[e.g.][]{bonnell98}, since binary orbits will suffer continual disruption as long as close encounters leading to mergers are occuring. If massive stars form primarily by collisions and not by accretion, the calculations we present in this paper will not apply.

Given a seed binary, in this paper we calculate how pre-main sequence stellar evolution and mass transfer will modify its properties. In \S~\ref{parameters}, we determine the range of parameters for which mass transfer can occur, and in \S~\ref{outcome} we discuss the outcome of mass transfer when it does occur. We discuss the implications of this process for the massive binary population, and related points, in \S~\ref{discussion}, and we summarize our conclusions in \S~\ref{summary}.

\section{The Properties of Mass Transfer Binaries}
\label{parameters}

The problem of binary pre-stellar binary evolution depends on both the accretion histories of the stars and on the properties of their orbit. Even if we limit ourselves to protostars accreting at constant rates and moving in circular orbits of constant size, there are four parameters: the accretion rates onto each of the two stars,  the offset in time between formation of the two stars, and the semi-major axis of the orbit. Since fully exploring this parameter space would be time consuming and not particularly informative, we further simplify it by assuming that the two binary components form coevally or nearly so, reducing the space to three dimensions. Our goal is simply to map out for what ranges of accretion rate and orbital separation mass transfer is a possibility. Our assumption that the two components of the binary are roughly coeval and that the larger component has a higher time-averaged accretion rate (necessarily the case for coeval, constant accretion rate systems) is supported by recent simulations of massive star formation \citep{krumholz07a}. 

The first step in exploring this parameter space is to determine the protostellar mass-radius relation as a function of accretion rate, which we do in \S~\ref{massradius}. Using these tracks, in \S~\ref{rlofsection} we determine the minimum semi-major axis required for Roche lobe overflow (RLOF) to occur in a binary system consisting of two stars of mass $M_1$ and $M_2$, $M_1>M_2$, with secondary-to-primary mass ratio $q=M_2/M_1\le1$, orbiting with semi-major axis $a$. The total system mass is $M$, and the binary accretes at a rate $\dot{M}_{\rm acc}$. Under our assumption of coeval formation, the ratio of accretion rates onto the two components is $\dot{M}_{\rm 2,acc}/\dot{M}_{\rm 1,acc}=M_2/M_1=q$, where the subscript ``acc" indicates the change in mass due to accretion into the system, rather than transfer between the two stars.

\subsection{The Protostellar Mass-Radius Relation}
\label{massradius}

%\begin{figure}
%\plotone{trackplot}
%\caption{\label{trackplot}
%Radius versus mass for constant accretion rates of $10^{-5}-10^{-2}$ $\msun$ yr$^{-1}$ (\textit{solid lines}), computed using the simple one-zone \citet{mckee03} model. For comparison we also show the results of the detailed numerical calculation of \citet{palla91, palla92} for accretion rates of $10^{-5}$ $\msun$ yr$^{-1}$ and $10^{-4}$ $\msun$ yr$^{-1}$ (\textit{dot-dashed line}). The sharp rise in $R$ that all models show results from the start of shell deuterium burning.\\
%}
%\end{figure}

%%%%%%%%%%%%%%%%%%%%%%%%%%%%%%%%%%%%%%%%%%%%%%%%%%%%%%%%%%%%%%%%%%%%%%%%%%%%
%\begin{figure}
%\centerline{\psfig{file=trackplot.eps,width=9.5cm}}
%\plotone{trackplot}
%\caption{Radius versus mass for constant accretion rates of $10^{-5}-10^{-2}$ $\msun$ yr$^{-1}$ (\textit{solid lines}), computed using the simple one-zone \citet{mckee03} model. For comparison we also show the results of the detailed numerical calculation of \citet{palla91, palla92} for accretion rates of $10^{-5}$ $\msun$ yr$^{-1}$ and $10^{-4}$ $\msun$ yr$^{-1}$ (\textit{dot-dashed line}). The sharp rise in $R$ that all models show results from the start of shell deuterium burning.\\
%\label{trackplot}}
%\end{figure}
%%%%%%%%%%%%%%%%%%%%%%%%%%%%%%%%%%%%%%%%%%%%%%%%%%%%%%%%%%%%%%%%%%%%%%%%%%%%%

For a given accretion rate onto a star we construct a track of radius versus mass using a simple protostellar model. To allow quick exploration of a large range of parameters, we use the one-zone model of \citet{mckee03} to generate our tracks. This model has been calibrated against the detailed numerical calculations of \citet{palla92}, and agrees to $\sim 10\%$. We refer readers to \citeauthor{mckee03} for a detailed description of the model, but here we summarize its most important features. In this model a protostar is assumed to be a polytropic sphere with a specified accretion rate as a function of time. At every time step of model evolution one uses conservation of energy to determine the new radius, including the effects of energy lost in dissociating and ionizing incoming gas, energy radiated away, and energy gained by Kelvin-Helmoltz contraction, deuterium burning, and the gravitational potential energy of accreting gas.

An accreting massive protostar passes through four distinct phases of evolution before reaching the zero-age main sequence (ZAMS). When the star first forms, it evolves passively without any nuclear burning. When its core becomes hot enough ($\sim 10^6$ K), the second phase begins: deuterium ignites, the star becomes fully convective and begins burning its accumulated deuterium reserve, and contraction of the core temporarily halts. Once the star exhausts its deuterium reserve, it enters a third phase in which it continues burning deuterium at the rate it is brought in by new accretion. In this phase the core resumes quasi-static contraction. As the core temperature continues to rise, its opacity drops, and eventually a layer forms that is stable against convection, initiating the fourth phase. Newly accreted deuterium cannot pass through the radiative layer to reach the core, and without an influx of new deuterium the core rapidly exhausts its supply and becomes convectively stable as well. Deuterium begins burning in a shell above the radiative zone, driving a rapid expansion of the star. As the core continues contracting, the radiative zone increases in size and the star resumes contraction. Eventually the core becomes hot enough to ignite hydrogen, and at that point the star joins the ZAMS.

\begin{figure}
%\centerline{\psfig{file=trackplot.eps,width=9.5cm}}
\plotone{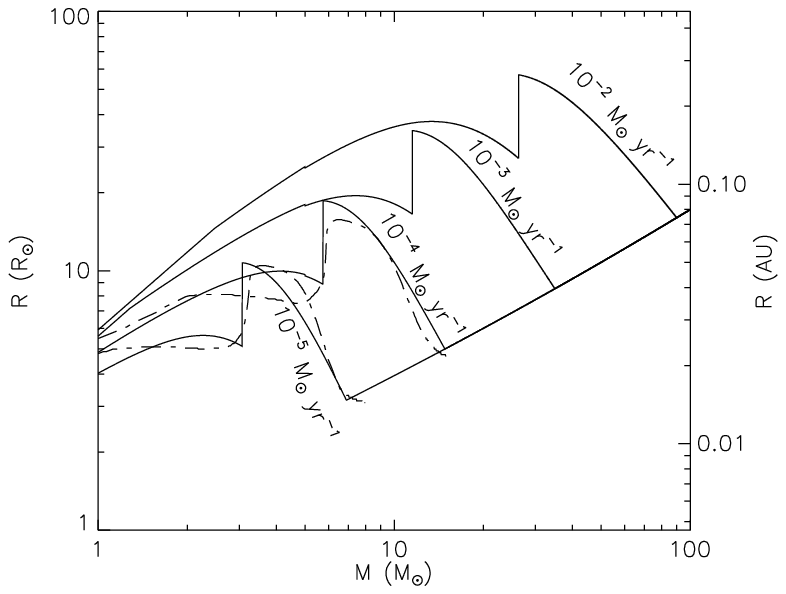}
%\plotone{f1.pdf}
\caption{Radius versus mass for constant accretion rates of $10^{-5}-10^{-2}$ $\msun$ yr$^{-1}$ (\textit{solid lines}), computed using the simple one-zone \citet{mckee03} model. For comparison we also show the results of the detailed numerical calculation of \citet{palla91, palla92} for accretion rates of $10^{-5}$ $\msun$ yr$^{-1}$ and $10^{-4}$ $\msun$ yr$^{-1}$ (\textit{dot-dashed line}). The sharp rise in $R$ that all models show results from the start of shell deuterium burning.\\
\label{trackplot}}
\end{figure}

Figure \ref{trackplot} shows some sample tracks of radius versus mass computed using this model. The sharp increase in radius that each model shows is the result of the start of deuterium shell burning. As the plots show, for the high accretion rates expected in massive star-forming regions, the radius can reach several tenths of an AU. We compute protostellar models for 600 values of $\dot{M}_{\rm acc}$, uniformly spaced in logarithm in the range $10^{-5}-10^{-2}$ $\msun$ yr$^{-1}$, accreting up to a maximum mass of 100 $\msun$. By this mass, stars have joined the ZAMS regardless of their accretion rate.

\subsection{Roche Lobe Overflow}
\label{rlofsection}

Having determined the protostellar mass-radius relation, we can now determine for what semi-major axes $a$ a binary of mass ratio $q$ (secondary-to-primary) will experience RLOF. The radius of the Roche lobe around star 1 is \citep{eggleton83}
\begin{equation}
\label{rocheradius}
R_{r1} \approx a \frac{0.49}{0.6 + q^{2/3} \log(1+q^{-1/3})};
\end{equation}
the Roche lobe radius around star 2 is given by an analogous expression with $q$ replaced by $q^{-1}$. For a given value of $q$ and $\dot{M}_{\rm 1,acc}$, the accretion rate onto the more massive star, it is straightforward to use the mass-radius tracks to compute whether RLOF ever occurs for a given value of $a$, and, if so, to determine the mass and state of each star when it does.

%\begin{figure}
%\plotone{rlofmap}
%\caption{\label{rlofmap}
%Minimum $a$ for RLOF (\textit{solid line}) as a function of accretion rate onto the more massive star $\dot{M}_{\rm 1,acc}$. The value of $q$ is indicated in each panel. Below the RLOF line, in the region of parameter space where overflow occurs, we show the mass $M_1$ (in solar units) of the donor star at overflow (\textit{shaded regions}). Binaries with $a$ below the dashed line experience overflow before the primary starts deuterium shell burning, while those with larger values of $a$ experience overflow after the onset of shell burning.\\ 
%\vspace{0.2in}
%}
%\end{figure}

%%%%%%%%%%%%%%%%%%%%%%%%%%%%%%%%%%%%%%%%%%%%%%%%%%%%%%%%%%%%%%%%%%%%%%%%%%%%
%\begin{figure}
%\centerline{\psfig{file=rlofmap.eps,width=9.2cm}}
%\plotone{rlofmap}
%\caption{Minimum $a$ for RLOF (\textit{solid line}) as a function of accretion rate onto the more massive star $\dot{M}_{\rm 1,acc}$. The value of $q$ is indicated in each panel. Below the RLOF line, in the region of parameter space where overflow occurs, we show the mass $M_1$ (in solar units) of the donor star at overflow (\textit{shaded regions}). Binaries with $a$ below the dashed line experience overflow before the primary starts deuterium shell burning, while those with larger values of $a$ experience overflow after the onset of shell burning.\\ 
%\label{rlofmap}}
%\end{figure}
%%%%%%%%%%%%%%%%%%%%%%%%%%%%%%%%%%%%%%%%%%%%%%%%%%%%%%%%%%%%%%%%%%%%%%%%%%%%%

\begin{figure}
%\centerline{\psfig{file=rlofmap.eps,width=9.2cm}}
\plotone{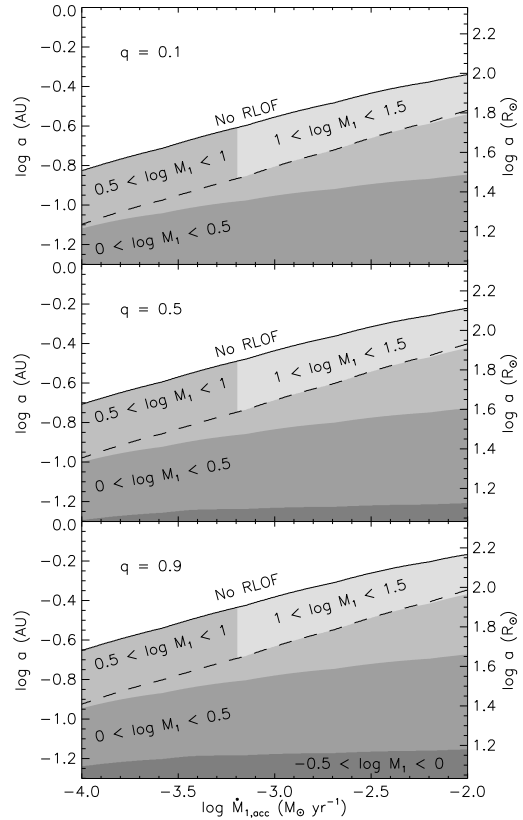}
%\plotone{f2.pdf}
\caption{Minimum $a$ for RLOF (\textit{solid line}) as a function of accretion rate onto the more massive star $\dot{M}_{\rm 1,acc}$. The value of $q$ is indicated in each panel. Below the RLOF line, in the region of parameter space where overflow occurs, we show the mass $M_1$ (in solar units) of the donor star at overflow (\textit{shaded regions}). Binaries with $a$ below the dashed line experience overflow before the primary starts deuterium shell burning, while those with larger values of $a$ experience overflow after the onset of shell burning.\\
\label{rlofmap}}
\end{figure}

Figure \ref{rlofmap} summarizes the results for 3 different values of $q$. We find that in rapidly accreting systems, RLOF will occur when the primary star is $M_1 \gtsim 10$ $\msun$ and the semi-major axis $a$ is a few tenths of an AU or less; a rough analytic estimate for the critical value of $a$ below which RLOF occurs is 
\begin{equation}
a_{\rm crit}\approx 0.2\,\,\dot{M}_{-4}^{1/4}(1+q)^{1/3}\,\,{\rm AU},
\label{acrit}
\end{equation}
where $\dot{M}_{-4}=\dot{M}_{\rm 1,acc}/10^{-4}$\,M$_\odot$ yr$^{-1}$.
Thus, in the context of the model proposed here, observed systems such as WR20a, D33J013346.2+304439.9, and the eclipsing binaries in the SMC, with masses $>20$ $\msun$ and separations $\ltsim 0.25$ AU, are well within the expected overflow range for reasonable accretion rates and initial mass ratios. We further find that it is always the more massive star that overflows its Roche lobe, rather than the less massive one. This is because, despite the larger Roche radius around the more massive star, its higher accretion rate leads this star to have a larger radius before the onset of shell burning, and to start deuterium shell burning sooner. Finally, note that much of the parameter space allowed for transfer, the region between the dashed and solid lines in Figure \ref{rlofmap}, is due to the shell burning phase, which roughly doubles the radius of the star and therefore doubles the minimum value of $a$ for which transfer can occur.

If we were to relax our assumption that the stars are coeval, our conclusion would be strengthened, because simulations indicate that in massive binary systems the more massive companion generally forms earlier \citep{krumholz07a}. Since the mass at which deuterium shell burning starts increases with accretion rate (see Fig.~\ref{trackplot}), and the protostellar radius at any time increases with both mass and accretion rate, this general result should be robust.

\section{The Outcome of Roche Lobe Overflow}
\label{outcome}

\subsection{Time Scales}

Before we attempt to calculate the outcome of RLOF, it is helpful to review some relevant timescales for the problem. The shortest is the orbital period, $P=3.7 a_{0.1}^{3/2} M_{10}^{-1/2}$ days, where $a_{0.1}$ is the semi-major axis in units of $0.1$ AU and $M_{10}$ is the total system mass in units of 10 $\msun$. This is the time scale on which mass lost from the primary star will reach the vicinity of the secondary; actual accretion may take longer if the gas has to be processed through a disk. Next is the dynamical, sound-crossing time of the donor star. The sound speed varies greatly from the stellar core to the surface, so to be conservative and obtain the longest possible time scale we evaluate this using the surface sound speed, which gives $\tau_{\rm dyn} = R_{*1}/c_{\rm s,surf} \sim 31R_{50} T_{4}^{-1/2}$ days, where $R_{50}$ is the stellar radius in units of $50$ $\rsun$ and $T_{4}$ is the stellar surface temperature in units of $10^4$ K. The timescale $\tau_{\rm dyn}$ describes the time required for the star to adjust mechanically as it loses mass. Finally, there is the Kelvin-Helmholtz time of the donor star, $\tau_{\rm KH} \sim GM_1^2/(R_{*1} L_1) \sim 6200 M_{10}^2 R_{50}^{-1} L_{4}^{-1}$ yr, where $L_{4}$ is the star's luminosity in units of $10^4$ $\lsun$. This describes the time required for the star to adjust thermally to mass loss.

The most important point to take from this calculation is that $\tau_{\rm KH}$ is by far the largest timescale in the problem, so that stars will be unable to adjust their thermal (as opposed to mechanical) structure to mass transfer that occurs on orbital or dynamical timescales. Thus, we may approximate stars as behaving adiabatically during mass transfer. A corollary of this is that, as gas in the donor star expands adiabatically in response to mass loss, deuterium burning will slow dramatically, since the burning rate is extremely temperature sensitive. Thus, once the star overflows its Roche lobe, its envelope will not have its specific entropy altered by further nuclear burning on a dynamical time scale.

\subsection{Stability of Mass Transfer}
\label{section:stability}

We first address the question of whether the RLOF leads to stable or unstable mass transfer. Transfer is stable if adiabatic mass loss shrinks the star at a rate faster than the Roche lobe around it shrinks, and is unstable otherwise. If the transfer is unstable, the donor star can change its mass by order unity on a time scale $\tau_{\rm dyn}$, and transfer will stop only when a new mechanical equilibrium is established. 

To check stability, we must evaluate how mass transfer changes both the Roche radius and the stellar radius. The former is easy to compute. If no mass is lost from the system, then conservation of angular momentum demands that the radius of the system shrink at a rate
\begin{equation}
\frac{\dot{a}}{a}=-2\frac{\dot{M}_1}{M_1} \left(1-\frac{1}{q_0}\right),
\end{equation}
where $q_0$ is the secondary-to-primary mass ratio at the time when mass transfer begins and $\dot{M}_1$ is the rate of change of the primary's mass due to mass transfer (as opposed to accretion onto the system). Since $\dot{M}_1<0$ and $q_0<1$, this means that $\dot{a}<0$; the semi-major axis of the orbit shrinks in response to mass transfer. From equation (\ref{rocheradius}), the Roche radius around the primary star varies approximately as
\begin{eqnarray}
\frac{\dot{R}_{r1}}{R_{r1}} & \approx & \frac{\dot{M}_1}{M_1} (1-q_0) \cdot
\nonumber \\
& &
\quad
\left[\frac{2}{q_0} + \frac{2 \log (1+q_0^{-1/3}) - (1+q_0^{1/3})^{-1}}{1.8 q_0^{1/3} + 3 q_0 \log (1+q_0^{-1/3})}\right].
\end{eqnarray}
The expression in square brackets is always positive for $q_0<1$, so the Roche radius around the donor star shrinks as well.

For the range of parameter space we explore, a star undergoing RLOF does so before reaching the ZAMS, and at a mass below $30$ $\msun$. Thus, the star is gas pressure-dominated, at least in its outer layers. If RLOF occurs before the onset of deuterium shell burning, the star is convective throughout, and thus is well-described as an isentropic $n=3/2$ polytrope. Since a star with this structure expands in response to mass loss at constant entropy (see below), we learn immediately that RLOF will be unstable in this case, because the star expands as the Roche radius contracts. Thus, systems whose parameters fall below the dashed lines in Figure \ref{rlofmap} are always unstable.

If RLOF occurs after deuterium shell burning starts, the star consists of a lower specific entropy inner radiative zone and a significantly larger, high specific entropy envelope. Shell burning eventually raises the temperature in the envelope to the point where it becomes radiative as well \citep{palla91}, but at the onset of shell burning it is still convective. RLOF occurs at the start of shell burning, since this is the point of maximum radius, so the envelope will also be an isentropic $n=3/2$ polytrope, with a different specific entropy than the interior radiative zone.

Following the analysis of \citet{paczynski72} for a red giant, we can approximate this configuration in terms of a ``centrally condensed" polytrope, a polytropic sphere with a point mass in its center. This is a less accurate approximation than it is for a red giant, since deuterium shell burning in a protostar only expands the envelope by a factor of slightly more than $2$, as opposed to more than an order of magnitude for a red giant. However, we can regard the pure polytropic model of a star before shell burning and the centrally condensed polytropic model for a star during shell burning as limiting cases.

Centrally condensed $n=3/2$ polytropes obey a mass-radius relation \citep{paczynski72}
\begin{equation}
\label{conpolytrope}
R= K E^{2/3} M^{-1/3},
\end{equation}
where $K$ is a constant that depends on the specific entropy in the convective envelope and $E$ is a dimensionless number that depends only on $\xi$, the ratio of the core mass to the total mass. Since the star behaves adiabatically as mass is removed, $K$ remains constant. If the mass of the radiative core is also unchanged, a reasonable approximation when transfer begins, then the radius of the star varies as
\begin{equation}
\label{stabilitycondition}
\frac{\dot{R}_{*1}}{R_{*1}} = -\frac{\dot{M}_1}{3 M_1} \left\{1+2\left[\frac{\xi_0}{E(\xi_0)}\right]\left.\frac{dE}{d\xi}\right|_{\xi=\xi_0}\right\},
\end{equation}
where $\xi_0$ is the core mass ratio at the start of mass transfer. The star expands for $\xi_0<0.22$, and shrinks for $\xi_0>0.22$.

\begin{figure}
%\centerline{\psfig{file=q0min.eps,width=9.5cm}}
\plotone{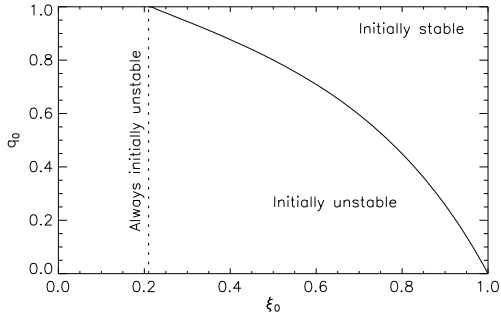}
%\plotone{f3.pdf}
\caption{Minimum initial mass ratio $q_0$ required to make mass transfer stable for a given initial core mass fraction $\xi_0$. The dotted line at $\xi_0=0.22$ indicates the minimum $\xi_0$ below which mass transfer is always unstable regardless of the value of $q_0$.\\
\label{q0min}}
\end{figure}

Since $R_{r1} = R_{*1}$ at the start of mass loss, mass transfer is unstable unless $\dot{R}_{*1}<\dot{R}_{r1}$. This condition is met only if
\begin{eqnarray}
\lefteqn{\frac{1}{3}\left\{1+2\left[\frac{\xi_{0}}{E(\xi_{0})}\right]\left.\frac{dE}{d\xi}\right|_{\xi=\xi_{0}}\right\} < }
\nonumber \\
&& (1-q_0)\left[\frac{2}{q_0} + \frac{2 \log (1+q_0^{-1/3}) - (1+q_0^{1/3})^{-1}}{1.8 q_0^{1/3} + 3 q_0 \log (1+q_0^{-1/3})}\right].
\end{eqnarray}
By evaluating $E$ and $dE/d\xi$ numerically \citep[e.g.\ table 1 of][]{paczynski72}, this condition lets us identify for each $\xi_0$ what values of the initial mass ratio $q_0$ produce stability. Figure \ref{q0min} shows that for $\xi_0<0.22$ no stability is possible because the star expands rather than contracting in response to mass loss.

%\begin{figure}
%\plotone{q0min}
%\caption{\label{q0min}
%Minimum initial mass ratio $q_0$ required to make mass transfer stable for a given initial core mass fraction $\xi_0$. The dotted line at $\xi_0=0.22$ indicates the minimum $\xi_0$ below which mass transfer is always unstable regardless of the value of $q_0$.\\ 
%\vspace{0.2in}
%}
%\end{figure}

%%%%%%%%%%%%%%%%%%%%%%%%%%%%%%%%%%%%%%%%%%%%%%%%%%%%%%%%%%%%%%%%%%%%%%%%%%%%
%\begin{figure}
%\centerline{\psfig{file=q0min.eps,width=9.5cm}}
%\plotone{q0min}
%\caption{Minimum initial mass ratio $q_0$ required to make mass transfer stable for a given initial core mass fraction $\xi_0$. The dotted line at $\xi_0=0.22$ indicates the minimum $\xi_0$ below which mass transfer is always unstable regardless of the value of $q_0$.\\ 
%\label{q0min}}
%\end{figure}
%%%%%%%%%%%%%%%%%%%%%%%%%%%%%%%%%%%%%%%%%%%%%%%%%%%%%%%%%%%%%%%%%%%%%%%%%%%%%

Stars without radiative cores have $\xi=0$, so if RLOF occurs before the onset of shell burning then $\xi_0=0$ and the transfer is unstable. Shell burning stars will have $\xi_0>0$, and they may be stable if $\xi_0$ and $q_0$ are large enough. The value of $\xi_0$ is determined by the size of the radiative core when it first appears, which is in turn set by the radius of the shell where the maximum luminosity that can be transported by radiation, $L_{\rm rad}(m)$, first rises to the point where it is equal to the luminosity passing through that shell, $L(m)$. Here, $m$ is the mass enclosed within a given shell, and varies from $m=0$ at the center of the star to $m=M_1$ at its surface.

From their numerical models, \citet{palla92} find that $L_{\rm rad}(m)=L(m)$ is first satisfied at $m/M_1=\xi_0=0.20$, $0.21$, and $0.41$ for accretion rates of $10^{-5}$, $3\times 10^{-5}$, and $10^{-4}$ $\msun$ yr$^{-1}$. We would like to extrapolate to higher accretion rates, but that is quite difficult because $\xi_0$ varies so non-linearly with $\dot{M}_{\rm 1,acc}$. Instead, we make some general observations to help understand the likely value of $\xi_0$. The non-linearity in the variation of $\xi_0$ with $\dot{M}_{\rm 1,acc}$ results from the complicated shape of $L_{\rm rad}(m)$ \citep[see figure 5 of][]{palla91}.
This shape admits two possible solutions to $L_{\rm rad}(m)=L(m)$, i.e.\ two mass shells where a convectively stable zone could appear, one at smaller $m$ and one at larger $m$. The reason $\xi_0$ is nearly unchanged between $\dot{M}_{\rm 1,acc}=10^{-5}$ and $3\times 10^{-5}$ $\msun$ yr$^{-1}$ is that $L_{\rm rad}(m)=L(m)$ is first satisfied at the same solution point in the two models; $\xi_0$ jumps between $\dot{M}_{\rm 1,acc}=3\times 10^{-5}$ and $10^{-4}$ $\msun$ yr$^{-1}$ because the point where $L_{\rm rad}(m)$ and $L(m)$ first become equal moves from the inner to the outer possible solution. Since the shape of the $L_{\rm rad}(m)$ curve does not suggest that there are other, larger mass solutions where $L_{\rm rad}(m)=L(m)$ might first occur, it seems unlikely that $\xi_0$ will be much larger than $0.41$. However, we caution that this is a tentative conclusion, and that one cannot confidently estimate $\xi_0$ without detailed numerical modeling. In particular, the value of $\xi_0$ in the models of \citet{palla91,palla92} appears to be sensitive to their choice of boundary condition.

If our tentative conclusion holds and $\xi_0\ltsim 0.5$ regardless of the accretion rate, then transfer will be unstable unless $q_0\gtsim 0.8$.

\subsection{The Fate of Stable Mass Transfer Binaries}
\label{stableoutcome}

Massive binary systems that initiate stable mass transfer lie in the region of parameter space above the line in Figure \ref{q0min}. Stable transfer is possible only if it starts because the more massive star has initiated shell burning, and possesses a radiative core. While transfer reduces the temperature in the envelope via adiabatic expansion and thus halts shell burning immediately, this seems likely to produce only a temporary reprieve. As the primary continues to contract toward the main sequence on a Kelvin-Helmoltz timescale, it cannot avoid re-heating and again igniting deuterium in its envelope, producing further expansion. Thus, there may be multiple episodes of shell burning, expansion, and stable mass transfer separated by time intervals of order $\tau_{\rm KH}$, each leading to the transfer of a relatively small amount of mass but over time accumulating to push the system considerably toward $q=1$. Determining the detailed evolution of a system undergoing this process requires the ability to follow the internal structure of the donor star as it repeatedly loses small amounts of gas from its envelope, cools adiabatically, and then heats up again. This is beyond the capabilities of our simple protostellar evolution model, and since our best estimate for $\xi_0$ indicates the unstable mass transfer is likely to be a more common occurrence than stable transfer, we do not pursue the problem further in this paper.

\subsection{Termination of Unstable Mass Transfer}

We have found that unless the mass ratio is already near unity and the donor star is shell burning, mass transfer is unstable and proceeds on a dynamical time scale. This makes it difficult to determine the precise outcome without a protostellar model that, unlike ours, is capable of following the accretion of very high entropy gas on this time scale. Numerical simulations of the transfer itself would also be helpful, since in the case of runaway transfer it is possible that mass lost from the primary may go into a circumbinary disk or a wind rather than accreting onto the secondary. Nonetheless, we can make some simple calculations to estimate how and whether mass transfer will cease.

Mass transfer will continue until either the primary shrinks within its Roche lobe, or until the secondary overflows its Roche lobe too. We discuss the latter outcome in \S~\ref{unstableoutcome}. We can check whether the former is a possibility by asking whether there exists a new mass ratio $q$, larger than the original mass ratio $q_0$, such that the primary and secondary are both within their Roche lobes. Since such an equilibrium must be established on the dynamical time scale, the donor star must be able to reach it adiabatically. To determine whether such an equilibrium exists, we first examine how the Roche radius and stellar radius for each star change with $q$ in \S~\ref{radevol}, and then we search for equilibria in \S~\ref{equilibria}.

\subsubsection{Roche Radius and Stellar Radius}
\label{radevol}

If the mass transfer conserves both total system mass and angular momentum, then the ratio of the semi-major axis $a$ to its value $a_0$ before transfer starts is
\begin{equation}
\label{aoverazero}
\frac{a}{a_0}= \left(\frac{q_0}{q}\right)^2\left(\frac{1+q}{1+q_0}\right)^4
\end{equation}
when the mass ratio reaches $q$. From equation (\ref{rocheradius}), the new Roche radius of the primary is related to the original one $R_{r1,0}$ by
\begin{equation}
\label{rrochechange}
\frac{R_{r1}}{R_{r1,0}} = \left(\frac{q_0}{q}\right)^2\left(\frac{1+q}{1+q_0}\right)^4
\left[\frac{0.6+q_0^{2/3} \log (1+q_0^{-1/3})}{0.6+q^{2/3} \log (1+q^{-1/3})}\right].
\end{equation}
The ratio of current to initial Roche radius for the second star,
$R_{r2}/R_{r2,0}$, is given by an equivalent expression with $q$ and
$q_0$ replaced with $q^{-1}$ and $q_0^{-1}$, respectively.  Note that once mass transfer on a dynamical timescale begins, the two stars cannot remain exactly synchronous, nor can they be treated as point masses, and thus equations (\ref{aoverazero}) and (\ref{rrochechange}) are approximate (J.~Goodman, 2007, private communication). 

Similarly, from the mass-radius relation (\ref{conpolytrope}), the
initial radius of star 1 and its radius once the mass ratio reaches $q$ are related by
\begin{equation}
\label{rstarchange}
\frac{R_{*1}}{R_{*1,0}} \approx \left(\frac{1+q}{1+q_0}\right)^{1/3} 
\left[\frac{E\left(\frac{1+q}{1+q_0} \xi_0\right)}{E(\xi_0)}\right]^{2/3}.
\end{equation}
This relation will break down if $\xi=\xi_0(1+q)/(1+q_0)$ becomes too
large, because formally a centrally condensed polytrope goes to zero
radius as $\xi\rightarrow 1$. In reality, the radiative core has a
finite radius, which the simulations of \citet{palla91,palla92} show
is $\sim 1/2$ of a dex smaller than the total stellar radius at the
onset of shell burning. However, we find below that
the radius of the primary star can never become this small before the
secondary overflows its Roche lobe, so equation (\ref{rstarchange}) is
an adequate approximation for our purposes.

Now consider the secondary star. Mass transfered from the primary to
the secondary will not have time to cool and will therefore retain its
specific entropy; it may even gain entropy due to an accretion shock
as it falls onto the secondary. Since this high entropy material is
being transferred to the smaller potential well of the second star, it
will rapidly expand the secondary's radius, and eventually cause it to
overflow its Roche lobe as well.

We can obtain a rough lower bound on the secondary's radius by
treating it as a centrally condensed polytrope as well. In this
approximation, we take the secondary's radius at the onset of transfer
to be negligible in comparison to its radius after mass transfer
starts, and thus treat the secondary at the start of transfer as a
point mass. After an amount of mass $\Delta M$ has been
transfered, the secondary has mass $M_2=M_{2,0}+\Delta M$ and
envelope-to-core mass ratio
$\xi_{2}=\Delta M/M_{2,0}$. Since the specific entropy of the gas that
is transferred is the same as or larger than it was before the transfer, the
constant $K$ for the secondary is at a minimum the same as for the primary at the
onset of transfer,
\begin{equation}
K=R_{*1,0} M_{1,0}^{1/3} E(\xi_{0})^{-2/3} = 
R_{r1,0} M_{1,0}^{1/3} E(\xi_{0})^{-2/3}.
\end{equation}
The radius of the secondary is therefore bounded below by
\begin{equation}
\label{rstar2}
R_{*2} > R_{*1,0} \left[\frac{1+q}{(1+q_0)q}\right]^{1/3}
\left\{\frac{ 
E\left[\frac{q_0}{q}\left(\frac{1+q}{1+q_0}\right)\right]
}{E(\xi_0)}
\right\}
\end{equation}
at the point when the mass ratio is $q$.

\subsubsection{Existence of Mechanical Equilibria}
\label{equilibria}

\begin{figure}
%\centerline{\psfig{file=radcomparison.eps,width=9.2cm}}
%\epsscale{0.6}
\plotone{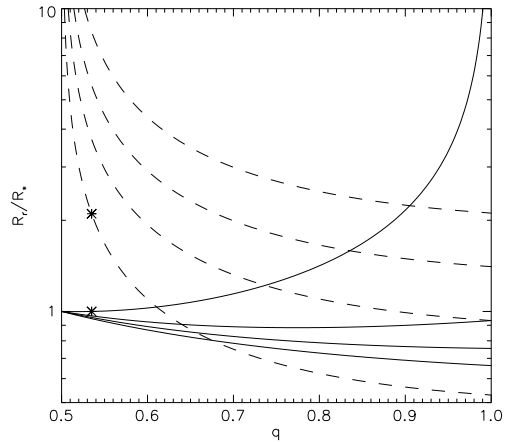}
%\plotone{f4.pdf}
%\epsscale{1.0}
\caption{\label{radcomparison}
Ratio of Roche radius $R_r$ to stellar radius $R_*$ for the primary
star (\textit{solid lines}) and the secondary star (\textit{dashed
lines}) for a system with $q_0=0.5$ initially. We show curves for
$\xi_0=0, 0.25, 0.5,$ and $0.75$, from lowest to highest. A system
stabilizes if there is a value of $q$ for which $R_r/R_* > 1$ for both
stars. The asterisks mark an example of such a stable point: a system
with $q_0=0.5$, $\xi_0=0.75$ stabilizes at $q=0.54$.
}
\end{figure}

Having determined how the Roche radius and stellar radius for each
star vary as the stars exchange mass and $q$ changes, we are now in
a position to determine whether there is a state the system can reach
adiabatically in which neither star is overflowing its Roche
lobe. Such an equilibrium exists if and only if there is a value
$q>q_0$ such that $R_{*1}\leq R_{r1}$ and $R_{*2} \leq R_{r2}$. We
also require $q\leq 1$. Since the envelope of the primary has higher
entropy than that of the secondary, the secondary will certainly
overflow its Roche lobe by the time $q=1$ is reached, even though our
lower limit in equation (\ref{rstar2}) does not reflect this.
We illustrate an example of the problem graphically in Figure
\ref{radcomparison}, which shows the ratio of Roche radius to stellar
radius for both stars as a function of $q$ for $q_0=0.5$ and various
values of $\xi_0$.

\begin{figure}
%\centerline{\psfig{file=qstable.eps,width=9.2cm}}
\plotone{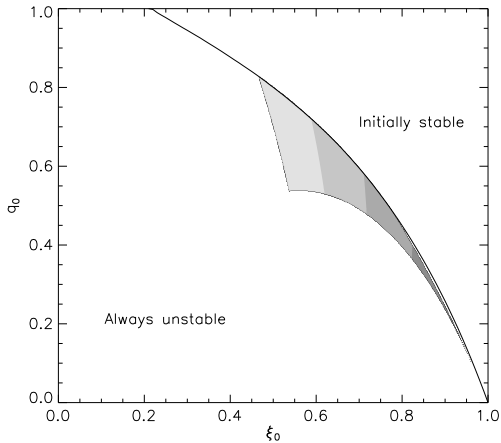}
%\plotone{f5.pdf}
\caption{Value of $q$ for which mass transfer terminates, as a
function of $\xi_0$ and $q_0$. Outside the shaded region, mass
transfer is either always stable (\textit{top right region}), or there
is no value of $q$ for which it will terminate (\textit{bottom left
region}). Inside the shaded region, the grayscale shows the value of
$q$ for which mass transfer terminates. The lightest region
corresponds to $q>0.8$, the next darkest region to $0.6<q\leq 0.8$,
the next to $0.4<q\leq 0.6$, and so forth.  Although $\xi_0$
is uncertain, in \S\ref{section:stability} we argue on the
basis of the models of \citet{palla91, palla92} that 
it is likely to be $\lesssim0.5$.
\label{qstable}}
\end{figure}

As the figure shows, for a given value of $q_0$ and $\xi_0$ there may
or may not be a value of $q$ satisfying the condition that both stars
are within their Roche radii. If there is such a $q$, then mass
transfer will halt once the system reaches that value.
We summarize these results in Figure \ref{qstable}, which shows the
value of $q$ for which a system stabilizes as a function of $q_0$ and
$\xi_0$. In the upper right white region of the plot, systems are
initially stable and mass transfer does not occur on a dynamical time
scale (see \S~\ref{stableoutcome}). In the lower left region, systems are always unstable, and
there is no value of $q$ for which the the donor star can shrink
within its Roche lobe. Systems whose parameters fall within the shaded
region in between are initially unstable, but can stabilize and stop
transferring mass for some value of $q$. Generically, we find that
systems are unable to reach equilibrium unless they fall within a
relatively small region of parameter space. Note that even this
certainly overestimates the size of the stable region, since our
estimate for $R_{*2}$ is only a lower limit. Thus, once transfer
starts, it generally continues until both stars overflow their Roche
lobes.

\subsection{The Fate of Unstable Mass Transfer Binaries}
\label{unstableoutcome}

What is the fate of a mass-transfer binary in which the primary cannot
shrink inside its Roche lobe before the secondary also overflows its
Roche lobe? Such a system will become a contact binary, or possibly
even a common envelope system if the semi-major axis decreases
enough. Equation (\ref{aoverazero}) implies that in a binary that
transfers enough mass to reach $q=1$, the semi-major axis shrinks by a
factor of $2^4 q_0^2 /(1+q_0)^4$. For $q_0=0.1$, this is a factor of
10, and almost certainly produces common envelope evolution, or
possibly even a merger. However, for $q_0=0.5$, the change in
separation is only 20\%, so a contact binary seems the more likely
outcome.

Once a contact binary or common envelope forms, over a thermal time
the envelopes of stars will equilibrate to the same specific
entropy. If the primary has a radiative core it will not equilibrate,
but the radiative core may well disappear temporarily because as it
expands and cools
adiabatically, its opacity will increase, and it may transition back
to convective instability. If this happens, it will reach the same
specific entropy as the two envelopes. The result will be an
entirely isentropic contact binary. Since the two stars are
isentropic, gas will be equally strongly bound to each of the two
potential wells, and the system will approach equal masses. As the
stars radiate they will continue contracting, their envelopes will
shrink, and eventually contact will cease before the stars reach the
ZAMS. The result will be a main sequence detached binary with a mass
ratio close to unity.

Note that this behavior is different than that commonly observed in
main sequence or post-main sequence contact binaries, which do not
approach equal masses. In those systems, the two stars are
chemically different (for example one may have consumed all of its
hydrogen), or one or both may contain degenerate cores that will not
reach the same specific entropy as the envelopes. It is these
differences that produce unequal masses.

After contact ceases and the stars contract to the ZAMS, the orbit
will circularize and the binary will become synchronous.  The 
characteristic timescale for the former has been estimated by 
\citet{goodman98a} (their eq.~15).  Taking representative
parameters, the circularization timescale
for an equal mass binary with separation $a \sim0.1$\,AU, and
constituent masses of $10$ M$_\odot$ and radii $R\sim0.02$\,AU,
is $t_{\rm circ}\sim10^2$\,yr $(R/0.02\,{\rm AU})^{3/2}[(a/R)/5]^{21/2}$.  
The very strong dependence on the ratio $a/R$ makes this estimate 
quite uncertain: a factor of two change in $a/R$ increases $t_{\rm circ}$
by a factor of $10^3$.  Importantly, however, the timescale is 
shorter than or comparable to the characteristic thermal time.
The synchronization timescale ($t_{\rm syn}$) can be estimated 
from the work of  \citet{goldreich89}.  Using the same 
parameters as for the estimation of $t_{\rm circ}$, we find a
synchronization timescale of $t_{\rm syn}\sim t_{\rm circ}$
with again a similar and very strong dependence on the magnitude 
of $a/R$.  Although uncertain, these estimates indicate that 
close massive binaries shold be synchronous and on circular orbits
relatively early in their main sequence lifetimes.

\section{Discussion}
\label{discussion}

A primary result of this work is that there is a critical $a$ below
which RLOF occurs (eq.~\ref{acrit}; Figure \ref{rlofmap}).  Thus, for
massive binaries that form at a separation of a few tenths of an AU,
RLOF and mass transfer occur before either star reaches the ZAMS.  For
a relatively wide range of parameter space, mass transfer is unstable
and we argue that the binary mass ratio will be driven toward equality
on a dynamical timescale.  For high system accretion rates, the ratio
of the radiative core mass to the envelope mass of the primary
protostar ($\xi_0$) can be $\sim 0.5$, and in some cases the system
can stabilize for some values of $q<1$. Somewhat unequal mass binaries
are the result when the stars reach the ZAMS (Figure \ref{qstable}).

\subsection{``Twin" Binaries}

A particularly curious feature of the massive binary population is the high proportion of systems with mass ratios very close to unity. The mass ratio of WR20a is $q=0.99\pm 0.05$ \citep{bonanos04, rauw05}, and that of D33J013346.2+304439.9 is $q=0.90\pm 0.15$ \citep{bonanos06}. Thirteen of the 21 detached binaries in the \citet{hilditch05} sample have $q>0.85$, and five have best fit values of $q$ consistent with $q=1.0$ \citep{pinsonneault06}. Due to its small size, the statistical significance of this sample is unclear \citep{lucy06}, but a priori it seems quite unusual that almost 25\% of the eclipsing binaries in the SMC with orbital periods shorter than 5 days should have mass ratios consistent with $1.0$. Nor does it seem likely a priori that the most massive binary known should have a mass ratio within 5\% of unity. 

Accretion from a circumbinary disk, as in the model \cite{bate00}, provides a way of producing an anti-correlation between semi-major axis and binary mass ratio, such as is observed even for low mass stars \citep{tokovinin00}. However, to produce mass ratios near unity, this mechanism requires that binary protostars initially form with very small masses and then accrete many times their original mass from a circumbinary disk. While such low mass seeds may form in the cold, lower density cores from which low mass stars form, both analytic work \citep{krumholz06b} and simulations \citep{krumholz07a} suggest that radiation feedback largely prevents the formation of small ``seed" stars in massive protostellar cores. In the simulations of \citet{krumholz07a}, secondaries may form by fragmentation out of a massive unstable disk (with subsequent migration), or via capture of a fragment formed elsewhere in the core, but in either case the primary at the time of binary formation is at least a few $\msun$ in mass, and the secondary at least a few tenths of $\msun$. Such a system likely could not reach a mass ratio of unity via the \citeauthor{bate00} mechanism. However, as we have shown here, mass transfer between massive accreting protostars provides a natural, and for close binaries inevitable, mechanism for attaining mass ratios near unity. That said, the circumbinary accretion mechanism of \citeauthor{bate00} probably operates in tandem with mass transfer --- particularly at lower masses --- bringing binaries closer together and pushing mass ratios to larger values, before mass transfer occurs and produces $q$ of nearly $1.0$ for tight binaries.

\subsection{Observational Predictions}

These calculations lead to definite predictions about the properties of massive binaries, which will be directly testable against future samples of binary systems larger than those available today. Since there is both a critical mass and a critical semi-major axis required for RLOF to occur, we predict that true twins, systems with mass ratios $q>0.95$, should be significantly more common among stars with masses $\gtsim 5-10$ $\msun$ with semi-major axes $\ltsim 0.25$ AU than among a binary population with either lower mass or larger separation. These twins should be in nearly zero eccentricity orbits, since mass transfer and evolution into a contact binary will circularize orbits rapidly.

Furthermore, systems with small values of $q_0$ that begin mass transfer will experience large reductions in the system semi-major axis. For example, a system with $q_0 = 0.1$ will reduce its separation by a factor of 10, likely producing a merger. This may lead to a deficit of low mass companions to massive stars at separations significantly smaller than a tenth of an AU. Unfortunately, predictions regarding close binaries with small $q$ are very hard to test observationally because such systems are difficult to detect.

Massive stars should generally have larger mass ratios in clusters with high surface density, since these likely formed from higher pressure gas and thus produced larger accretion rates onto the stars within them \citep{mckee03}. However, this effect is likely to be rather weak, since the critical semi-major axis only varies with the quarter power of accretion rate, and very high accretion rates make it easier for stars to remain unequal mass after the onset of transfer because they decrease the fraction of the star's mass that goes into the extended envelope. As a result, there may be fewer true twins in very high surface density systems, even if there are more massive stars with mass ratios $\gtsim 0.5$. In any event, due to weak dependence on the accretion rate, we consider this possibility less promising than searching for correlations of twin fraction with mass and semi-major axis.

\subsection{Primordial Stars}

One final note is that, although we have not discussed primordial stars, the mechanism we have discussed may well operate in them too. The critical ingredients for mass transfer are a phase of deuterium shell burning to produce large radii, and a close companion onto which to transfer mass. The binary properties of primordial stars are completely unknown, so we cannot comment on whether the second condition is likely to be met. However, a phase of deuterium shell burning does seem likely. Primordial stars probably form at high accretion rates \citep{tan04a}, giving them large radii \citep{stahler86}. In present-day stars, a radiative barrier forms because opacity in a stellar interior decreases with temperature. The opacity source is primarily free-free transitions of electrons, and the availability of electrons for this process does not depend strongly on metallicity in an ionized stellar interior. Thus, primordial stars likely form radiative barriers much like present day ones, and undergo deuterium shell burning.

\section{Summary}
\label{summary}

Massive, rapidly accreting protostars can reach radii of tenths of an AU during their pre-main sequence evolution, largely because they undergo a phase of deuterium shell burning that swells their radii. During this evolutionary phase, massive protostars with close companions will overflow their Roche lobes and transfer mass. Such transfer is always from the more massive, rapidly accreting star to the smaller one, since radius increases with both mass and accretion rate. Using simple protostellar structure and evolution models, we evaluate the range of separations and mass ratios for which mass transfer is expected to occur, and compute the likely outcome of the transfer.

We find that, for the expected accretion rates in mass star-forming regions, binaries at separations of several tenths of an AU or closer will undergo mass transfer, with some dependence on the exact accretion rate and the initial mass ratio. This process always pushes binaries toward mass ratios of unity. For some accretion rates and initial mass ratios, mass transfer will either be stable initially, or will terminate on its own before reaching $q=1$. For many systems, though, it will only halt when the two stars form a contact binary. The stellar envelopes in such systems will rapidly reach almost equal masses and specific entropies, and then the stars contract onto the main sequence, forming a ZAMS binary with a mass ratio very close to unity.

Pre-main sequence mass transfer represents a heretofore unknown phase of binary star formation and evolution, one that has likely affected a significant fraction of massive spectroscopic binaries. It provides a natural explanation for the puzzling phenomenon of massive twins, high mass binaries with mass ratios that are consistent with $q=1.0$. We predict based on this finding that twins should be significantly more common among stars $\gtsim 5-10$ $\msun$ in mass at separations $\ltsim 0.25$ AU than among either less massive or more distant stars, and that mass ratios should generally increase with binary mass and decrease with separation. The weak dependence of $a_{\rm crit}$ in equation (\ref{acrit}) on the system mass accretion rate and the initial mass ratio suggests that this result should not depend significantly on formation environment. Surveys such as OGLE that detect statistically large samples of massive binaries are rapidly making these predictions testable.

\acknowledgements We thank J. Goodman, S.~W. Stahler, and J.~M. Stone for discussions and helpful comments on the manuscript, and B. Paczy\'nski for prompting us to think about this problem. MRK acknowledges support from NASA through Hubble Fellowship grant \#HSF-HF-01186 awarded by the Space Telescope Science Institute, which is operated by the Association of Universities for Research in Astronomy, Inc., for NASA, under contract NAS 5-26555. TAT acknowledges support from a Lyman Spitzer Jr.\ Fellowship.

%\bibliographystyle{apj}
%\bibliography{refs}

\end{document}